\documentclass[12pt,a4paper]{article}
\usepackage{amsmath}
\usepackage{times}
\usepackage[dvips]{graphicx}
\input epsf
\date{}
\begin{document}
\author{{S.M. Troshin, N.E. Tyurin}\\
\small\it{Institute for High Energy Physics},\\
\small\it{Protvino, Moscow Region, 142281 Russia}}
\title
{{Limiting energy dependence of  spin parameters in small-angle elastic pp-scattering}}
\maketitle
\begin{abstract}
We consider limiting behavior with energy of spin parameters at
small values of t close to zero which corresponds to the
saturation of the Froissart-Martin bound for the total cross
sections.
\end{abstract}

\section*{Introduction}
The most global characteristic of the hadronic collision is the
total cross--section and the
nature of the total--cross section rising energy dependence
is an open problem in  hadron physics at large distances.
There is a widespread  conjecture  that the Froissart--Martin bound
should be saturated at high energies and there are several model parameterizations for the total
cross-sections  using $\ln^2 s$ dependence for $\sigma_{tot}(s)$ even at available energies.
Saturation of the Froissart--Martin bound results from the corresponding saturation of unitarity
for the partial amplitudes $|f_l(s)|=1$ at $l\leq \sqrt{s}\ln s$ and it leads to the prediction
\cite{aub} of limiting behaviour of other observables, e.g. $\sigma_{el}(s)\sim\ln^2 s$, slope of
 diffraction cone $B(s)\sim\ln^2 s$, ratio of real to the imaginary part of the forward scattering
 amplitude $\rho(s)\sim \ln^{-1}(s)$. The above limiting energy dependencies are related
 to observables averaged over spin degrees of freedom.
It is evident that the mentioned regularities should are related
somehow  to the nonperturbative QCD.

It is important nowadays  to have  corresponding limiting energy
dependencies  for spin parameters; at RHIC the experiments are
performed with polarized proton beams of high energies in the
 region $\sqrt{s}=200-500$ GeV and those dependencies can be useful for estimation of the magnitude
 of spin parameters and the likelihood of the assumptions on the role of the helicity amplitudes
  and possibility to neglect some of them.

  Discussion of a role and magnitude
  of helicity-flip amplitudes in small-angle elastic scattering
has a long history  and is an important issue in
the studies of the spin properties of high-energy diffraction. Recent results of the $pp2pp$ experiment
at RHIC on $A_N$ measurements at $|t|$ in the region of 0.01 to 0.03 $(GeV/c)^2$ have
 indicated that hadronic spin-flip amplitude is non-zero and the
 dominating high energy scattering mechanism, Pomeron, can flip the helicity \cite{pp2pp}.
 Mechanism of Pomeron helicity flip based on the two-pion contribution to the scattering
 amplitude has been proposed in \cite{kane} and in \cite{usp} it was shown that unitarity generates
 phase difference between helicity flip and nonflip Pomeron amplitudes, so there is no need to introduce C-odd
 exchange called Odderon to generate nonzero asymmetry $A_N$. It was predicted \cite{usp} that $A_N$ at
 small t values decreases like $1/\ln s$. The bounds for the helicity amplitudes
at $t=0$ have been discussed in \cite{mah} and recent discussion
of these bounds and models for spin dependence in high
 energy proton scattering has been given in \cite{but1}.

 This brief note is devoted
 to the discussion of the same problem: it
  considers limiting energy dependencies of spin parameters which
  result from the saturation of unitarity for the helicity amplitudes.

We need to consider all helicity amplitudes and do not neglect
some of them from the very beginning. For instance the double
helicity-flip
 amplitudes can also contribute
into $A_N$ and  their behavior at high energies is also
 important  for the spin correlation
parameters and the total cross-section differences in experiments with two
polarized beams available at RHIC nowadays. It could also affect the small $t$ dependence of the
unpolarized differential cross-sections at the LHC.

\section{Unitarity bounds and limiting energy dependence}
Unitarity bounds for the helicity amplitudes have been obtained in
\cite{flip}. One should recall that saturation of unitarity for
the helicity amplitudes  leads to    a peripheral dependence of
the amplitudes $f_i(s,b)$ $(i=2,4,5)$ on the impact parameter $b$
at high energy, i.e.
\[
|f_i(s,b=0)|\rightarrow 0
\]
at $s\rightarrow\infty$. At small impact parameters only helicity
non-flip amplitudes survive. This is a consequence of the explicit
 unitarity
representation for the helicity amplitudes in the rational form.
This fact allows one to get better bounds for the helicity-flip
amplitudes at $t=0$ compared to those obtained in \cite{mah,but1}.

In what follows we will consider  limit  $s\to\infty$ and $-t$ is
small. Assuming saturation of the
 unitarity bounds obtained in \cite{flip}, we will have the following limiting behaviour
 of the helicity amplitudes
\[
F_1(s,t),\, F_3(s,t),\, \sim s\ln^2 s,\;F_5(s,t)\sim
\sqrt{-t}s\ln^2 s,
\]
\begin{equation}\label{lim}
F_2(s,t)\sim s\ln s,\,\,F_4(s,t)\sim -t s\ln^3 s.
\end{equation}

The  expressions for spin observables in terms of helicity
amplitudes in $pp$ elastic scattering have the following form. The
analyzing power $A_N$   looks like
\begin{equation}
\sigma A_N=-\mbox{Im}[(F_1+F_2+F_3-F_4)F_5^*],\label{1.66}
\end{equation}
where
\[
\sigma =\frac{1}{2}(|F_1|^2+|F_2|^2+|F_3|^2+|F_4|^2+4|F_5|^2)
\]
stands for the  differential  cross--section (up  to  the
normalization factor).
The expressions for the initial state spin correlation parameters are given
in terms of the $pp$--scattering helicity amplitudes as follows:
\begin{eqnarray}
\sigma A_{LL}  & = & \frac{1}{2}(-|F_1|^2-|F_2|^2+|F_3|^2+|F_4|^2),\nonumber \\[1ex]
\sigma A_{NN} & = & \mbox{Re}(F_1F_2^*-F_3F_4^*)+2|F_5|^2, \nonumber\\[1ex]
\sigma A_{SS} & = & \mbox{Re}(F_1F_2^*+F_3F_4^*), \nonumber \\[1ex]
\sigma A_{SL} & = & \mbox{Re}[(-F_1+F_3+F_2+F_4)F_5^*].\label{1.69}
\end{eqnarray}

In the experiments with polarized beams, besides the spin
correlation parameters which are differential characteristics
of the scattering processes there are possibilities to study
 global characteristics such as  differences of the total cross
sections for  pure spin states of the initial particles.
These quantities $\Delta \sigma _L(s)$ and $\Delta \sigma _T(s)$ correspond to definite
orientations of  spins of the initial particles in the
longitudinal
and transverse directions respectively.
They are defined as follows (where arrows indicate the spin directions
for particles $a$ and $b$ in the process $a+b\to X$):
\[
\Delta \sigma _L(s)= \sigma^{tot}_{^{\rightarrow}_{\leftarrow}}(s)-
\sigma^{tot}_{^{\rightarrow}_{\rightarrow}}(s),
\]
\begin{equation}
\Delta \sigma _T(s)=\sigma^{tot}_{\uparrow\downarrow}(s)- \sigma ^{tot}_{\uparrow\uparrow}(s).
\label{1.68}
\end{equation}
Here the first or top arrow refers to particle $a$.
For the case of $pp$ scattering according to optical theorem
the quantities  $\Delta \sigma _L(s)$ and $\Delta \sigma _T(s)$
are determined by the values of the helicity
amplitudes at $t=0$:
$\mbox{Im}[F_1(s,t=0)-F_3(s,t=0)]$ and $-\mbox{Im}F_2(s,t=0)$ respectively.

Using limiting dependencies Eqs. (\ref{lim}) and above formulas for observables in terms
of helicity amplitudes we arrive to the following energy dependencies of spin observables
in the limit of high energies and small and fixed values of $t$:
\begin{equation}
\label{limspin}
A_N(s,t), \, A_{NN}(s,t), \, A_{SS}(s,t), \, A_{SL}(s,t)\sim 1/\ln s,
\end{equation}
while
\begin{equation}
\label{all}
A_{LL}(s,t)\to 1.
\end{equation}
It should be noted that the limiting behavior of $A_N$
coincides with the model result of \cite{usp} and decreases with energy
as $1/\ln s$ while the ratio of the single flip amplitude $F_5$ to non-spin flip
amplitudes $F_1$ and $F_3$ is constant in the high energy limit.

In the high energy limit the following relations should take place in the
above kinematical limit:
\begin{equation}\label{eq}
A_{NN}(s,t)\simeq -A_{SS}(s,t)
\end{equation}
and
\begin{equation}\label{eq1}
A_{LL}(s,t)\gg A_{NN}(s,t), \,  A_{SS}(s,t), \, A_{SL}.
\end{equation}
To obtain conclusive result on the difference of the total cross-sections with longitudinal spin
orientations the additional model assumptions are needed, since it is the difference of the two
amplitudes with the same energy dependence.
Otherwise, the difference of total cross-sections with transverse spin orientations
has an unambiguous increasing limiting energy dependence
\begin{equation}\label{dl}
\Delta \sigma _T(s)\sim \ln s .
\end{equation}
The magnitude of the helicity amplitude $F_2$ at small $-t$ could be not small, it increases
with energy and it possibly can be measured directly
at RHIC through the measurements of $\Delta\sigma_T$.
This measurement would provide an
important contribution to the studies of the spin properties of diffraction.

The knowledge of limiting dependence
$\Delta\sigma _T(s)$, Eq. (\ref{dl}), allows one to obtain
 transverse spin structure function $h_1(x)$
in the limit of small Bjorken $x$ since in the low-$x$ region structure function $h_1$
is related to  discontinuity of the helicity
amplitude $F_2$ of the  quark-hadron forward
scattering, i.e.
\begin{equation}\label{h1}
xh_1(x)\sim \ln \frac{1}{x}.
\end{equation}
This is an important estimate which show that the structure function
$h_1$ can be large at small $x$ and it enhances the case  of the
transversity experimental studies   in Drell-Yan processes in  polarized
proton-proton scattering.

Finally, we would like to note that at the LHC energies one could expect  noticeable
effects in the small $t$ differential cross-section due to the contribution of the $F_4$
amplitude which is proportional to $s\ln^3 s$. Due to very high energy of the LHC this
effect could take place in the region of small $t$ despite the fact that $F_4$ goes to zero as
$t$ in the limit $-t\to 0$. It might happen that contributions of the helicity flip amplitudes
$F_2$, $F_4$ and $F_5$ would affect the results on the total cross-sections extraction performed
by the extrapolation to the optical point.

The limiting energy dependencies of spin parameters have been derived as a
result of the bounds saturation which follows from explicit solution
of unitarity for all helicity flip amplitudes and have the similar footing as
the saturation of the Froissart-Martin bound
for the total cross-section averaged over spins of the initial particles.
Thus, to be selfconsistent one should assume saturation of the above unitary bounds for spin
parameters too if the Froissart-Martin bound is supposed to be saturated
at high energies since the underlying mechanism in both cases is the same, i.e. saturation
of unitarity bounds for partial amplitudes (helicity flip and non flip ones). The respective
dependencies are given by Eqs. (\ref{limspin},\ref{eq}-{\ref{h1}) and they can be relevant for the
experimental studies of elastic scattering at RHIC and LHC.

\end{document}